\documentclass[10pt,runningheads]{llncs}

\usepackage{epsfig}
\usepackage{verbatim}
\usepackage{url}
\usepackage[colorlinks=true,urlcolor=blue]{hyperref}

\usepackage{adjustbox}
\usepackage{multirow}
\usepackage[utf8]{inputenc}
\usepackage{textcomp}
\usepackage{pgfplots}
\usepgfplotslibrary{groupplots}
\pgfplotsset{width=10cm,compat=1.9}
\usepackage{caption}

\def\SA{\mbox{\rm {\sf SA}}}

\def\T{\mbox{\rm {\sf T}}}
\def\P{\mbox{\rm {\sf P}}}

\newcommand{\occ}{\ensuremath{{\mit occ}}}

\def\+{\!+\!}
\def\-{\!-\!}

\begin{document}
\title{Fast Indexes for Gapped Pattern Matching\thanks{This research is
    supported by Academy of Finland through grant 319454.}}

\author{
Manuel C{\'a}ceres\inst{1}
\and
Simon J. Puglisi\inst{2}
\and
Bella Zhukova\inst{2}
}

\institute{
    Department of Computer Science, University of Chile\\
    Santiago, Chile\\
    \email{mcaceres@dcc.uchile.cl}    
\and
    Department of Computer Science, University of Helsinki\\
    Helsinki Institute for Information Technology (HIIT)\\
    Helsinki, Finland\\
    \email{\{puglisi,bzhukova\}@cs.helsinki.fi}\\[1ex]
}

\date{}

\maketitle

\begin{abstract}
We describe indexes for searching large data sets for variable-length-gapped (VLG) patterns. VLG patterns are composed of two or more subpatterns, between each adjacent pair of which is a gap-constraint specifying upper and lower bounds on the distance allowed between subpatterns. VLG patterns have numerous applications in computational biology (motif search), information retrieval (e.g., for language models, snippet generation, machine translation) and capture a useful subclass of the regular expressions commonly used in practice for searching source code. Our best approach provides search speeds several times faster than prior art across a broad range of patterns and texts.
\end{abstract}

\section{Introduction}

In the classic pattern matching problem, we are given a string $\P$ (the pattern or query) and asked 
to report all the positions where it occures in another (longer) string $\T$ (the text).
This problem has been very heavily studied and has applications throughout computer science.

In this paper we consider a variant on the classic pattern matching problem, called variable length gap 
(VLG) pattern matching. In VLG matching, the query $\P$ is not a single string but is composed of $k \ge 2$ 
strings (subpatterns) that must occur in order in the text. Between each subpattern, a number of characters 
may be allowed to occur, an upper and lower bound on which is specified as part of the query. Formally, our 
problem is as follows.

\begin{definition}[Variable Length Gap (VLG) Pattern Matching~\cite{BGVW12}]
Let $\T$ be a string of $n$ symbols drawn from alphabet $\Sigma$ and
$\P$ be a pattern consisting of $k \ge 2$ subpatterns (i.e. strings)
$p_0, \ldots, p_{k-1}$, each consisting of symbols also drawn from 
$\Sigma$, and having lengths $m_0, \ldots, m_{k-1}$, and $k-1$ gap
constraints $C_0, \ldots, C_{k-2}$, such that $C_i = \langle \delta_i, \Delta_i\rangle$
with $0 \le \delta_i \le \Delta_i < n$ specifies the smallest ($\delta_i$)
and largest ($\Delta_i$) allowable distance between a match of $p_i$ and
$p_{i+1}$ in $\T$. Find all matches --- reported as $k$-tuples $i_0,\ldots, i_{k-1}$
where $i_j$ is the starting position for subpattern $p_j$ in $\T$ --- such
that all gap constraints are satisfied.
\end{definition}

In computational biology, VLG matching is used in the discovery of and search for {\em motifs} --- 
i.e. conserved features --- in 
sets of DNA and protein sequences (see, e.g.,~\cite{MPVZ05,P14}). For example, the following is a protein 
motif from the rice genome (see~\cite{MPVZ05}) expressed as a VLG pattern:

$$\mbox{{\tt MT}[115, 136]{\tt MTNTAYGG}[121, 151]{\tt GTNGAYGAY}.}$$

A similar motif concept in music information retrieval means VLG matching also finds applications in mining 
and searching for characteristic melodies~\cite{CIR98} and other musical structures~\cite{CIMRTT02,FG08} in 
sequences of musical notes expressed in chromatic or diatonic notation. Bader et al.~\cite{BGP16} point out 
several more applications of VLG matching in information retrieval and related fields
such as natural language processing (NLP) and machine translation. For example, Metzler and Croft~\cite{MC05} define 
a language model in which query terms occuring within as certain window of each other must be found (in NLP, such 
terms are said to be {\em colocated}). Locating tight windows of a document containing the set of words contained in a 
search engine query is the problem of query-biased snippet generation~\cite{TTHW07}. In machine translation, VLG 
matching is used to derive rule sets from text collections to boost effectiveness of automated 
translation systems~\cite{L07}. 


VLG matching has a big parameter space and it is easy to think of pathological combinations of pattern and 
text that lead to an exponential number of matches. Fortunately, however, in practice the problem gets naturally 
constrained in important ways. The gap constraints are always bound the length of documents under consideration, 
which in the case of source code or web pages means that usually $\delta_i$ and $\Delta_i$ (and so their difference) 
are in (at most) the tens-of-kilobytes range. In genomics and proteomics maximum gaps tend to be around 100 
characters or so (see, e.g.,~\cite{MPVZ05,P14}).  

\medskip

Because of the interesting and useful applications outlined above, VLG matching has received a great deal of attention in the past 20 years.
The vast majority of previous work deals with the {\em online} version of the problem in which both the pattern $\P$ and the text $\T$ 
are previously unseen and cannot be preprocessed~\cite{BFC08,BGVW12,BT10,FG08,HSSSS11,MPVZ05,SSSS15}. Our concern in this paper is the 
{\em offline} version of the problem, where $\T$ is known in advance and can be preprocessed and an index structure built and stored to 
later support fast search for previously unseen VLG patterns (the stream of which is assumed to be large, practically infinite). Almost
all work on the offline problem is of theoretical interest~\cite{L11,BG14}. The exception is the recent work of Bader, Gog, and Petri~\cite{BGP16}, 
who develop methods for the {\em offline} setting 
that use a combination of suffix 
arrays~\cite{MM93} and wavelet trees~\cite{GGV03,N13}. Bader et al. show that their index is an order of magnitude faster at VLG matching 
than are online methods, and several times faster than $q$-gram-based indexes, the likes of which were behind Google Code Search~\cite{C12}.


\paragraph{Contribution.} Our main contribution in the paper is to show that in practice, on a broad range of inputs typical in real 
applications of VLG matching, simple algorithms based on intersecting ranges of the suffix array corresponding to subpattern occurrences 
can be made very fast in practice, and comfortably outperform state-of-the-art methods based on wavelet trees.

We emphasise that none of our new approaches are particularly exotic. They are, however, very fast,
and so represent non-trivial baselines by which future (possibly more exotic) indexes for VLG pattern matching and related problems (such as
regex matching) can be meaningfully measured.


\paragraph{Roadmap.} The remainder of this paper is as follows.
Section~\ref{sec:sascan} then looks at a simple method for solving VLG matching that works by sorting and intersecting ranges of the suffix 
array that contain the occurrences of subpatterns of the VLG pattern. Sections~\ref{sec:filter} and~\ref{sec:textcheck} evolve this basic 
idea, presenting the results of small illustrative experiments along the way. In Section~\ref{sec:experiments} we compare our best performing 
method to the recent wavelet-tree-based approach of Bader et al., which represents the current state-of-the-art for indexed pattern matching 
(details of our test machine and data sets can also be found in Section~\ref{sec:experiments}). Reflections and directions for future work are 
then offered in Section~\ref{sec:conclusion}.

\section{VLG matching via sorting and scanning suffix array intervals}
\label{sec:sascan}

Essential to the methods for VLG matching we will consider in this and later sections is the {\em suffix array}~\cite{MM93}
data structure. The suffix array of $\T$, $|\T| = n$, denoted $\SA$, is an array $\mbox{SA}[0..n-1]$, which contains a 
permutation of the integers $0..n$ such that $\T[\SA[0]..n-1] < \T[\SA[1]..n-1] < \cdots <
\T[\SA[n]..n-1]$.  In other words, $\SA[j] = i$ iff $\T[i..n]$
is the $j^{\mbox{{\scriptsize th}}}$ suffix of $\T$ in ascending
lexicographical order.
Because of the lexicographic ordering, all 
the suffixes starting with a given substring $p$ of $\T$ form an interval $\SA[s..e]$, which can be determined by binary 
search in $O(|p|\log n)$ time. Clearly the integers in $\SA[s..e]$ correspond precisely to the distinct positions of 
occurrence of $p$ in $\T$ and once $s$ and $e$ are located it is straightforward to enumerate them in time $O(e-s)$.
\medskip

The starting point for our approaches is a baseline algorithm from the study by Bader et al. called {\sc SA-scan},
which makes use of the suffix array of $\T$. A pseudo-C++ fragment adapted from Bader et al.'s codebase capturing the 
main thrust of {\sc SA-scan} is shown in Fig.~\ref{fig:sascancode}. For ease of reading the code here assumes two subpatterns, but is easy to 
generalize for $k > 2$.

\begin{figure}[ht]
{\footnotesize {\tt
\begin{tabbing}
\hspace{0.5cm} \= \hspace{0.5cm} \= \hspace{0.5cm} \= \hspace{0.5cm} \= \hspace{0.5cm}\kill
\>       SA-scan(string\_type p1, string\_type p2, int min\_gap, int max\_gap)\string{\\
\> \>        //1:find intervals of SA containing subpattern occurrences\\
\> \>        std::pair$\mbox{{\tt <}}$int,int$\mbox{{\tt >}}$ interval1 = search(p1,T,SA);\\
\> \>        std::pair<int,int> interval2 = search(p2,T,SA);\\
\> \>        //2: copy positions of subpattern occurrence from SA and sort\\ 
\> \>        int m1 = interval1.second-interval1.first+1;\\
\> \>        int m2 = interval2.second-interval2.first+1;\\
\> \>        int *A = new int$\mbox{[m1]}$;\\
\> \>        int *B = new int$\mbox{[m2]}$;\\
\> \>        std::memcpy(A,SA+interval1.first,m1);\\
\> \>        std::memcpy(B,SA+interval2.first,m2);\\
\> \>        std::sort(A,A+m1);\\
\> \>        std::sort(B,B+m2);\\
\> \>        //3: intersect according to gap constraints\\
\> \>        for(int i=0,j=0; i<m1 \&\& j<m2; i++)\string{\\
\> \> \>          while(B[j] < (A[i] + min\_gap) \&\& j < m2) j++;\\
\> \> \>          while(j < m2 \&\& B[j] <= (A[i] + max\_gap))\string{\\
\> \> \> \>             result.push\_back(B$[$j$]$);\\
\> \> \> \>            j++;\\
\> \> \>           \string}\\
\> \>        \string}\\
\>     \string}
\end{tabbing}
}
}
\caption{A basic C++ implementation of the {\sc SA-scan} VLG matching algorithm suitable for $k=2$ subpatterns.}
  \label{fig:sascancode}
\end{figure}

The operation of {\sc SA-scan} can be summarized as follows. First, search for each of the $k$ subpatterns using $\SA$ 
to arrive at $k$ ranges of the $\SA$ containing the subpattern occurrences (in the code listing this is acheived by the 
two {\tt search} method calls). Next, for each range, allocate a memory buffer 
equal to the range's size and copy the contents of the range from $\SA$ to the newly allocated memory and sort the contents 
of the buffer (positions of subpattern occurrence) into ascending order. Finally, intersect the positions for subpatterns 
$p_0$ and $p_1$ with respect to the gap constraints. Experimenting with {\sc SA-scan} we observed the time taken to find 
the ranges of subpattern occurrences in $\SA$ constituted less than 1\% of the overall runtime, with the vast majority of 
time spent sorting.

Bader et al. use {\sc SA-scan} as a baseline from which to measure the success of their wavelet-tree-based method.
{\sc SA-scan} is natural enough, to be sure, but it does look suspiciously like a straw man. 
To start with, is {\tt std::sort} really the best we can do for sorting those arrays of integers? 
We replaced the {\tt std::sort} call with a call to an LSD radix sort of our own implementation (using a radix of 256) and 
replicated an experiment from Bader et al.'s paper, searching several text collections (including web data, source code, DNA, 
and proteins --- see Section~\ref{sec:experiments} for more details) 
for 20 VLG patterns ($k=2$, $\delta_i,\Delta_i = \langle 100,110\rangle$), composed of very frequent subpatterns drawn from the 200 most common
substrings of length 3 in each data set. 

\begin{figure}[thb]
  \begin{center}
  \includegraphics[width=0.9\linewidth]{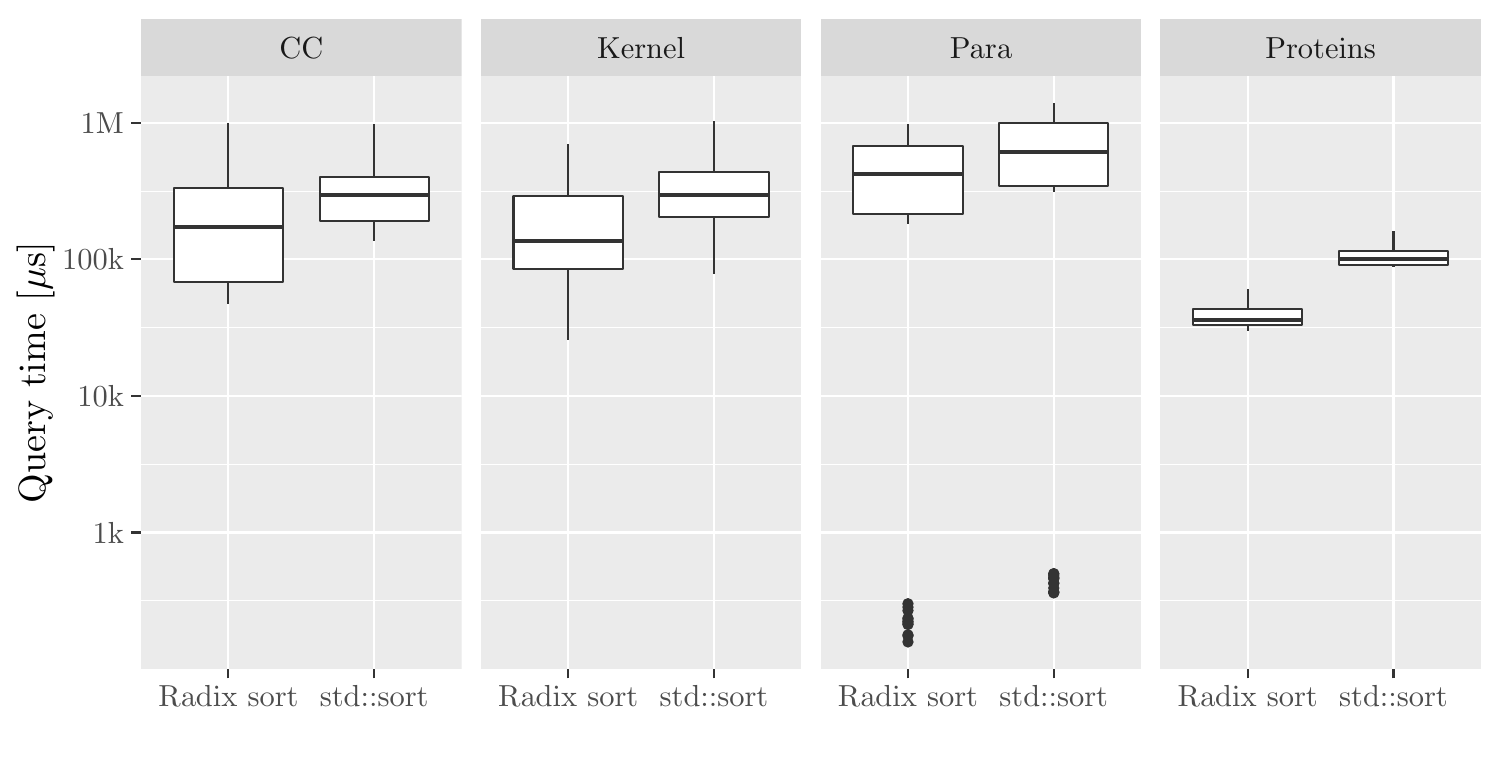}
  \caption{Time to search a 2GiB subset of the Common Crawl web collection (\protect\url{commoncrawl.org}).
for 20 VLG patterns ($k=2$, $\delta_i,\Delta_i = \langle 100,110\rangle$), composed of very frequent subpatterns drawn from the 200 most common
substrings of length 3 in the collection.
  }
  \label{fig:sascan}
  \end{center}
\end{figure}

Figure~\ref{fig:sascan} shows the results obtained on our test machine (see Section~\ref{sec:experiments} for specifications).
Using radix sort instead of {\tt std::sort}, {\sc SA-scan} becomes at least two times faster on the Kernel and Proteins 
datasets, almost twice as fast on CC, and more than 30\% faster on Para. A large if algorithmically-somewhat-unexciting leap forward --- but further improvements 
are possible\footnote{It is possible that further improvements from sorting alone are possible, using a more heavily engineered
sort function that our hand-rolled LSD radix sort. Our point here 
is that sorting is an important dimension along which {\sc SA-scan} can be optimized.}.

\section{Filter, filter, sort, scan}
\label{sec:filter}

Our first serious embellishment to {\sc SA-scan} aims to avoid sorting the full set of subpattern occurrences by filtering out 
some of the candidate positions that cannot possibly lead to matches. Specifically, we allocate a bitvector $F$ of $n/b$ bits 
initially all set to 0. We refer to $b$ as the {\em block size} of the filter. Logically, each bit represents a block of $b$ contiguous 
positions in the input text, with the $i$th bit corresponding to the positions $ib..i(b+1)-1$. In describing the use of the filter we 
assume two subpatterns $p_1$ and $p_2$ (with occurrences in $\SA[s_1..e_1]$ and $\SA[s_2..e_2]$, respectively), but the technique 
is easy to generalize for $k > 2$.

Having allocated $F$, we scan the interval $\SA[s_1..e_1]$ containing the occurrences of subpattern $p_1$ and for each element 
$i = SA[j]$ encountered, we set bits $F[(i+\delta)/b..(i+\Delta)/b]$ to $1$ to indicate that an occurrence of $p_2$ in any of 
the corresponding blocks of the input is a potential match. During the scan we also copy elements of the interval to an array 
$A_1$ of size $m_1 = e_1 - s_1 + 1$. We then scan the interval $\SA[s_2..e_2]$ containing the occurrences of the second subpattern 
and for each position $i$ encountered we check $F[i/b]$. If $F[i/b] = 0$ then $i$ cannot possibly be part of a match and can 
be discarded. Otherwise ($F[i/b] = 1$) we add $i$ to a vector $A_2$ of candidates. 
We then sort $A_1$ and $A_2$ and intersect them with respect 
to the gap constraints, the same as in the original {\sc SA-scan} algorithm. The hope is that $|A_2|$ is much less than 
$e_2 - s_2 + 1$, and so the time spent sorting prior to intersection will be reduced.  

There are two straightforward refinements to this approach. The first is to make the initial scan not necessarily over $\SA[s_1..e_1]$, 
but instead over the smaller of intervals $\SA[s_1..e_1]$ and $\SA[s_2..e_2]$. The only difference is that if the interval for $p_2$ 
(the second subpattern) happens to be smaller (i.e. $p_2$ has less occurrences in $\T$ than $p_1$) then we set bits $F[(i-\delta)/b..(i-\Delta)/b]$ 
(rather than $F[(i+\delta)/b..(i+\Delta)/b]$) to 1. Assuming $p_1$ is in fact more frequent than $p_2$, the second refinement is to 
perform a second round of filtering using the contents of $A_2$. More precisely, having obtained $A_2$, we clear $F$ (setting all bits 
to 0) and scan $A_2$ settings bits $F[(i-\delta)/b..(i-\Delta)/b]$ to 1 for each $i \in A_2$. We then scan $A_1$ and discard any element 
$i$ for which $F[i/b]$ now equals 0. Obviously it only makes sense to employ this heuristic if the initial filtering reduced the number 
of candidates, $|A_2|$, of the second subpattern significantly below $m_1$. In practice we found $m_2 < m_1/2$ led to a consistent speedup. 

Of course, these techniques generalize easily to $k>2$ subpatterns. The idea is that the output of the intersection of the first two 
subpatterns then becomes an input interval to be intersected with the third subpattern, and so on.

\begin{figure}[htb]
  \begin{center}
  \includegraphics[width=\linewidth]{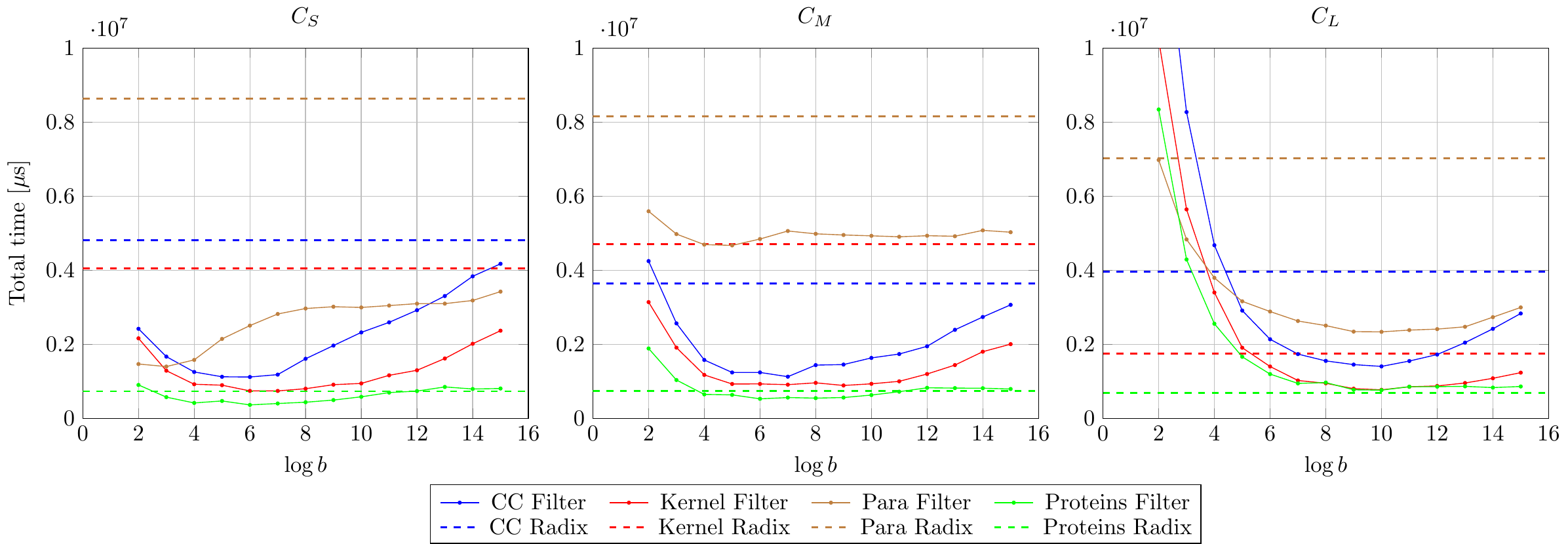}    
  \caption{Effect of filter granularity on search time. Ordinate is runtime in microseconds and mantissa is the logarithm (base 2) 
of the filter block size. Dashed lines show the time required by {\sc SA-scan} (using radix sort) without any filter. Each plot 
corresponds to a different set of 20 synthetically generated VLG patterns. All patterns contain 2 of the 200 most frequent subpatterns 
in each data set. We fixed the gap constraints $C_i = \langle \delta_i, \Delta_i\rangle$ between subpatterns to small 
($C_S = \langle 100,110 \rangle$), 
medium ($C_M = \langle 1000,1100 \rangle$), or large ($C_L = \langle 10000,11000 \rangle$). Section~\ref{sec:experiments} gives more 
details of data sets and pattern sets.}
  \label{fig:filter}
  \end{center}
\end{figure}

As Figure~\ref{fig:filter} shows, employing $F$ can reduce runtime immensely, but the improvement varies greatly with $b$. A good choice for 
$b$ depends on a number of factors. For each occurrence of $p_1$, we set $\lceil \frac{\Delta - \delta}{b} \rceil$ bits in $F$. 
Accesses to $F$ while setting these bits are essentially random (determined by the order of the positions of $p_1$, which are the 
lexicographic order of the corresponding suffixes of $\T$), and so it helps greatly if $b$ is chosen so that $F$, which has size $n/b$ 
bits, fits in cache. This can be seen in Figure~\ref{fig:filter}, particularly clearly for the CC, Kernel, and Protein data sets, where 
performance improves sharply with increasing $b$ until $F$ fits in cache (30MiB on our test machine) where it quickly stablizes (at 
$\log b = 3$ for CC and Kernel, and $\log b = 2$ for Protein). Runtimes then remain remain relatively fast and stable until $b$ becomes 
so large that the filter lacks specificity, from which point performance gradually degrades. Para has the same trend, though it is not 
immediately obvious --- because the data set is smaller (409MB) $F$ already fits in L3 cache when $b=2$. Section~\ref{sec:experiments} 
gives more details of data sets and pattern sets.

For the large-gap pattern set ($C_L$), where $\Delta - \delta = 1000$ the optimal choice of $b$ for all data sets is much higher --- $b = 1024$ 
in all cases ($b = 512$ has very similar performance). Here we are seeing the effect of the time needed to set bits in the filter. For example, 
for Kernel, $F$ already fits in L3 cache when $b = 8$, but at that setting $\lceil \frac{\Delta - \delta}{b} \rceil = 1000/8 = 125$ bits must 
be set in $F$ per occurrence of $p_1$. With $b = 1024$ or $512$, the number of bits set in $F$ per occurrence of $p_1$ is just 1 or 2, the same 
as it is at the optimal setting for the small-gap ($C_S$) and middle-gap ($C_M$) pattern sets. This effect can probably be largely alleviated
by employing two levels of filters or, alternatively, by implementing a method for setting a word of 1s at a time (effectively reducing
the time to set bits from $\lceil \frac{\Delta - \delta}{b} \rceil$ to $\lceil \frac{\Delta - \delta}{b\cdot w} \rceil$, where $w$ is the 
word size). 


\section{Direct text checking}
\label{sec:textcheck}

The filtering ideas described in the previous section can drastically reduce the amount of time spent per subpattern occurrence,
but the overall runtime is still $\Omega(\occ_1 + \occ_2)$, because both subpattern intervals are scanned in full. When the number 
of occurrences of the less frequent subpattern, say $p_1$, are significantly less than those of $p_2$, it is possible to get below 
that bound by scanning over only the occurrences of $p_1$, and for each occurrence, $i$, checking directly in the substring of text 
$\T[i+\delta..i+\Delta]$ for any occurrences of $p_2$, each of which corresponds to a match (or valid candidate match in the case 
$k>2$). If we use a linear time pattern matching algorithm such as that by Knuth, Morris, and Pratt~\cite{KMP} to search for the 
occurrences of $p_2$, runtime (for two subpatterns) becomes $\Theta(\occ_1\cdot(\Delta - \delta))$.

\begin{figure}[htb]
  \includegraphics[width=0.9\linewidth]{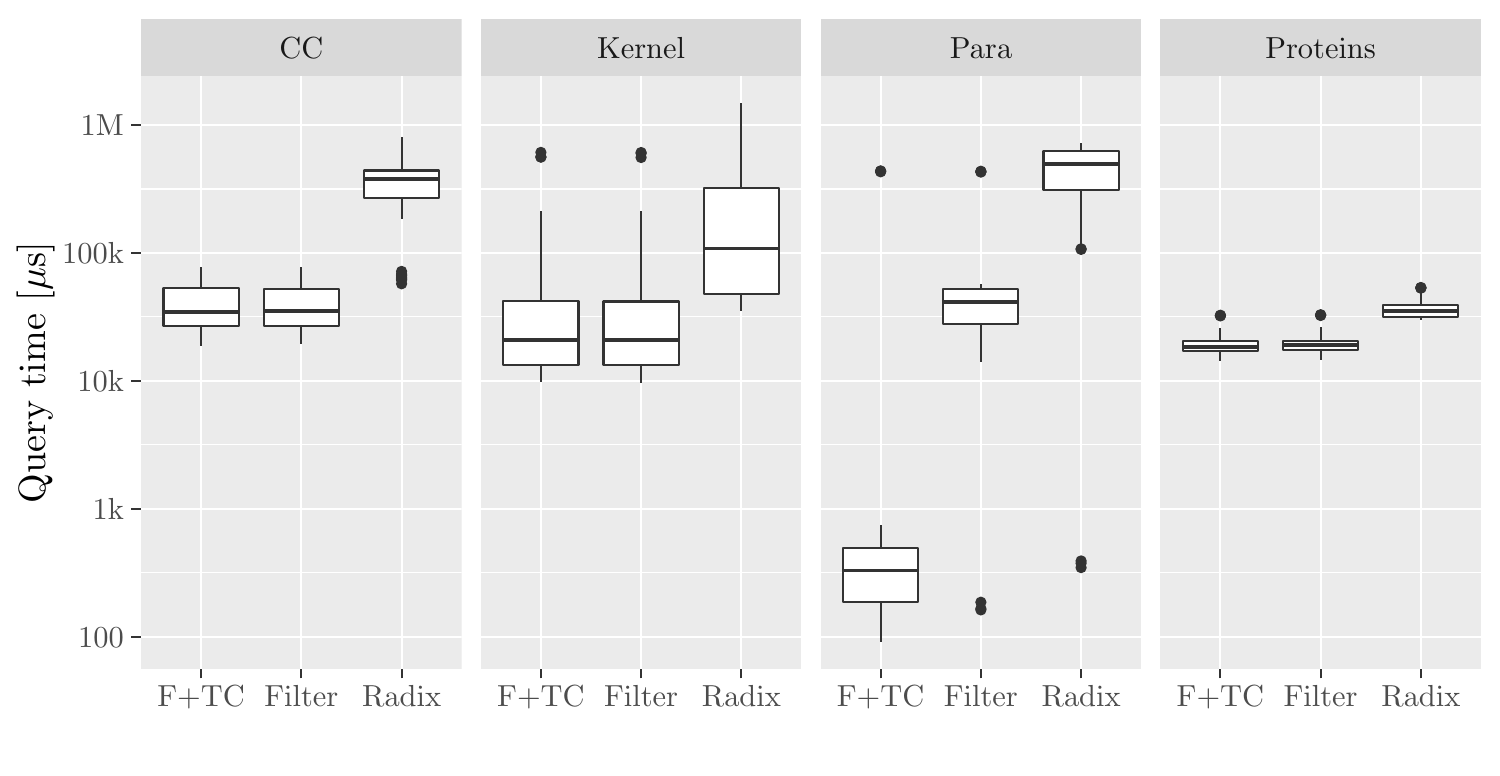}
  \caption{Direct text checking improves search times further ($k=2$).
  }
  \label{fig:textcheck}
\end{figure}

\begin{figure}[htb]
  \includegraphics[width=0.9\linewidth]{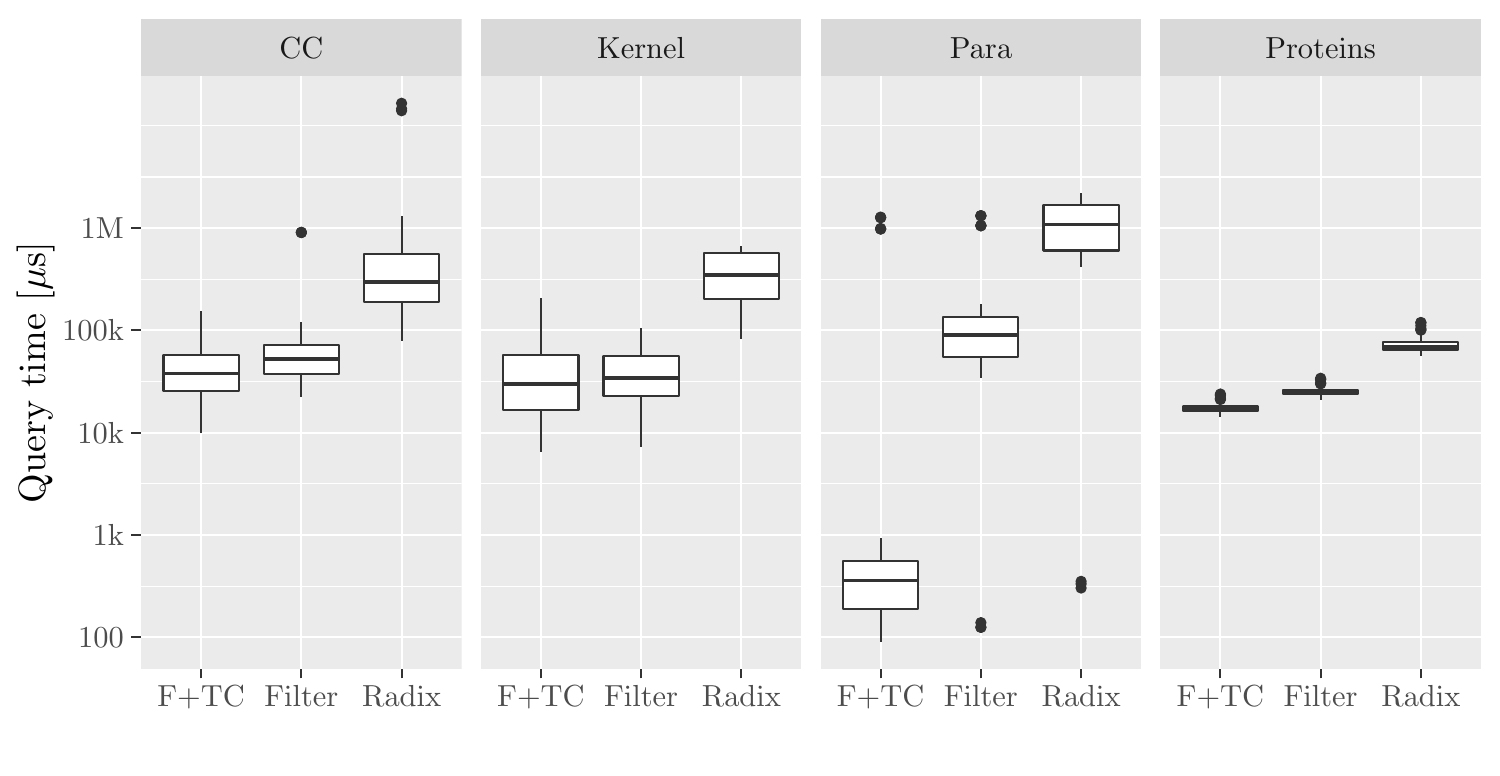}
  \caption{Direct text checking improves search times further ($k = 4$).
  }
  \label{fig:textcheck-k4}
\end{figure}

Employed by itself, this kind of text checking can lead to terrible performance when both $\occ_1$ and $\occ_2$ are large. However, 
when employed in concert with a filter, it can lead to significant performance gains, particularly in later rounds of intersection 
when $k>2$. Figure~\ref{fig:textcheck} illustrates this for $k=2$, along with the performance of the other versions 
of {\sc SA-scan} 
(Filter and Radix) 
we have decribed in previous sections. In sum, {\sc SA-scan} has been sped up by more than an order of 
magnitude on some data sets. In Fig.~\ref{fig:textcheck-k4} we see that the text checking 
heuristic makes an even bigger improvement when the number of subpatterns increases (from $k=2$ to $k=4$) because it 
is employed more often.

\section{Experimental Evaluation}
\label{sec:experiments}
In this section we compare the practical performance of our version of {\sc SA-scan} to the wavelet-tree-based method of
Bader et al., which is called {\sc WT}.
We use a variety of texts and patterns, which 
are detailed below (most of these have appeared in experiments described in previous sections). Our methodology in this 
section closely follows that of~\cite{BGP16}.

\paragraph{Test Machine and Environment.}
We used a 2.10\,GHz Intel Xeon E7-4830 v3 CPU equipped with
30\,MiB L3 cache and 1.5\,TiB of main memory. The machine had no other significant
CPU tasks running and only a single thread of execution was used.
The OS was Linux (Ubuntu 16.04, 64bit) running kernel 4.10.0-38-generic. Programs were
compiled using {\tt g++} version 5.4.0. 

\paragraph{Texts.} We use five datasets from different application domains:
\begin{itemize}
    \renewcommand\labelitemi{--}
\item CC is a 2 GiB prefix of a recent 145TiB web crawl from \href{http://commoncrawl.org}{commoncrawl.org}.
\item Kernel is a 2 GiB file consisting of source code of all (332) Linux kernel versions $2.2.X$, $2.4.X.Y$ and $2.6.X.Y$ downloaded from \href{http://kernel.org}{kernel.org}. The data set is very repetitive as only minor changes exist between subsequent versions.
\item Para is a 410 MiB, which contains 36 sequences of Saccharomyces Paradoxus, is provided by the \href{http://www.sanger.ac.uk/Teams/Team71/durbin/sgrp}{Saccharomyces Genome Resequencing Project}. There are four bases $\{A, C, G, T\}$, but  some characters denote an unknown choice among the four bases in which case $N$ is used.
\item Proteins is a 1.2 GiB sequence of newline-separated protein sequences (without descriptions, just the bare proteins) obtained from the \href{ftp://ftp.ebi.ac.uk/pub/databases/swissprot/release_compressed/uniprot_sprot.dat.gz}{Swissprot database}. Each of the 20 amino acids is coded as one letter.
\end{itemize}

\paragraph{Patterns.} As in~\cite{BGVW12}, patterns were generated synthetically for each data set. We fixed the gap constraints 
$C_i = \delta_i,\Delta_i$ between subpatterns to small ($C_S = \langle 100,110\rangle$), medium ($C_M = \langle 1000,1100\rangle$), or large ($C_L = \langle 10000,11000\rangle$).
VLG patterns were generated by extracting the 200 most common substrings of lengths 3, 5, and 7, which are 
then used as subpatterns. We then form 20 VLG patterns for each dataset, $k$ (i.e. number of subpatterns), and gap constraint by 
selecting from the set of 200 subpatterns. We emphasise that the generated patterns, while not specifically designed to be pathological, 
do represent relatively hard instances for {\sc SA-scan} because of the high frequency of each subpattern. 

\paragraph{Matching Performance for Different Gap Constraint Bands.}
Our first experiment aims to elucidate the impact of gap constraint size on query time. We fix the subpattern length $|p_i| = m_i = 3$.
Table~\ref{tab:gapcons} shows the results from VLG patterns consisting of $k = 2^1,\ldots,2^5$ subpatterns.
Our method, marked {\sc Filter+TC}, is always faster than {\sc WT}, with the exception of the large-gap $C_L$ pattern sets, where
on some data sets it yields to {\sc WT} (most likely due to the text-checking heuristic being less effective on $C_L$).
\begin{table}
\begin{center}

\let\center\empty
\let\endcenter\relax
\centering
\resizebox{1\width}{!}{




\centering
\begin{adjustbox}{width=\textwidth}

    \begin{tabular}{lllllllllllll}
    \hline
    \multicolumn{1}{l|}{\multirow{2}{*}{Method}} 
        & \multicolumn{3}{l|}{CC} 
        & \multicolumn{3}{l|}{Kernel} 
        & \multicolumn{3}{l|}{Para}
        & \multicolumn{3}{l}{Proteins}
        \\ 
    \cline{2-13} 
    \multicolumn{1}{r|}{} 
    & \multicolumn{1}{c|}{$C_S$} & \multicolumn{1}{c|}{$C_M$} & \multicolumn{1}{c|}{$C_L$} 
    & \multicolumn{1}{c|}{$C_S$} & \multicolumn{1}{c|}{$C_M$} & \multicolumn{1}{c|}{$C_L$} 
    & \multicolumn{1}{c|}{$C_S$} & \multicolumn{1}{c|}{$C_M$} & \multicolumn{1}{c|}{$C_L$}
    & \multicolumn{1}{c|}{$C_S$} & \multicolumn{1}{c|}{$C_M$} & \multicolumn{1}{c}{$C_L$} 
    \\ 
    \hline
    \multicolumn{13}{l}{k = 2}
    \\ 
    \hline
    \multicolumn{1}{l|}{Filter+TC}               
    & \multicolumn{1}{r|}{1391}      & \multicolumn{1}{r|}{1783}      & \multicolumn{1}{r|}{3329}      
    & \multicolumn{1}{r|}{744}      & \multicolumn{1}{r|}{1303}      & \multicolumn{1}{r|}{2243}  
    & \multicolumn{1}{r|}{2243}      & \multicolumn{1}{r|}{4170}      & \multicolumn{1}{r|}{2184} 
   & \multicolumn{1}{r|}{388}      & \multicolumn{1}{r|}{485}      & \multicolumn{1}{r}{976} 
    
    \\ \hline
    \multicolumn{1}{l|}{WT\_ALL}                 
    & \multicolumn{1}{r|}{17354}      & \multicolumn{1}{r|}{19736}      & \multicolumn{1}{r|}{21757}      
    & \multicolumn{1}{r|}{7393}      & \multicolumn{1}{r|}{10205}      & \multicolumn{1}{r|}{15891}      
    & \multicolumn{1}{r|}{12372}      & \multicolumn{1}{r|}{14683}      & \multicolumn{1}{r|}{3654}
    & \multicolumn{1}{r|}{8549}      & \multicolumn{1}{r|}{10547}      & \multicolumn{1}{r}{14470}
    
    \\ \hline
    \multicolumn{13}{l}{k = 4}           
    \\ \hline
    \multicolumn{1}{l|}{Filter+TC}               
    & \multicolumn{1}{r|}{1099}      & \multicolumn{1}{r|}{1739}      & \multicolumn{1}{r|}{3186}      
    & \multicolumn{1}{r|}{664}      & \multicolumn{1}{r|}{1158}      & \multicolumn{1}{r|}{1654}      
    & \multicolumn{1}{r|}{1214}      & \multicolumn{1}{r|}{8134}      &       \multicolumn{1}{r|}{9036}
    & \multicolumn{1}{r|}{370}      & \multicolumn{1}{r|}{580}      & \multicolumn{1}{r}{1609}
    \\ \hline
    \multicolumn{1}{l|}{WT\_ALL}                 
    & \multicolumn{1}{r|}{7452}      & \multicolumn{1}{r|}{7499}      & \multicolumn{1}{r|}{8770}      
    & \multicolumn{1}{r|}{2201}      & \multicolumn{1}{r|}{4810}      & \multicolumn{1}{r|}{2756}      
    & \multicolumn{1}{r|}{6554}      & \multicolumn{1}{r|}{9039}      & \multicolumn{1}{r|}{5230}
    & \multicolumn{1}{r|}{8486}      & \multicolumn{1}{r|}{9319}      & \multicolumn{1}{r}{12621}
    \\ \hline
    
    \multicolumn{13}{l}{k = 8}           
    \\ \hline
    \multicolumn{1}{l|}{Filter+TC}               
    & \multicolumn{1}{r|}{2185}      & \multicolumn{1}{r|}{1219} &      \multicolumn{1}{r|}{3057} 
    & \multicolumn{1}{r|}{328}      & \multicolumn{1}{r|}{733}      & \multicolumn{1}{r|}{3563}      
    & \multicolumn{1}{r|}{5}      & \multicolumn{1}{r|}{27}      & \multicolumn{1}{r|}{10300}      
    & \multicolumn{1}{r|}{354}      & \multicolumn{1}{r|}{581}      & \multicolumn{1}{r}{2231}
    \\ \hline
    \multicolumn{1}{l|}{WT\_ALL}
    & \multicolumn{1}{r|}{7919}      & \multicolumn{1}{r|}{6511}      & \multicolumn{1}{r|}{6054}      
    & \multicolumn{1}{r|}{1885}      & \multicolumn{1}{r|}{2260}      & \multicolumn{1}{r|}{1468}      
    & \multicolumn{1}{r|}{237}      & \multicolumn{1}{r|}{345}      & \multicolumn{1}{r|}{2518}      
    & \multicolumn{1}{r|}{11940}      & \multicolumn{1}{r|}{12867}      & \multicolumn{1}{r}{13450}
    \\ \hline
    
    \multicolumn{13}{l}{k = 16}           
    \\ \hline
    \multicolumn{1}{l|}{Filter+TC}               
    & \multicolumn{1}{r|}{674}      & \multicolumn{1}{r|}{930} &      \multicolumn{1}{r|}{3267} 
    & \multicolumn{1}{r|}{312}      & \multicolumn{1}{r|}{667}      & \multicolumn{1}{r|}{984}      
    & \multicolumn{1}{r|}{5}      & \multicolumn{1}{r|}{15}      & \multicolumn{1}{r|}{631}      
    & \multicolumn{1}{r|}{350}      & \multicolumn{1}{r|}{534}      & \multicolumn{1}{r}{2361}
    \\ \hline
    \multicolumn{1}{l|}{WT\_ALL}
    & \multicolumn{1}{r|}{7483}      & \multicolumn{1}{r|}{5580}      & \multicolumn{1}{r|}{7827}      
    & \multicolumn{1}{r|}{948}      & \multicolumn{1}{r|}{1497}      & \multicolumn{1}{r|}{1540}      
    & \multicolumn{1}{r|}{279}      & \multicolumn{1}{r|}{260}      & \multicolumn{1}{r|}{249}      
    & \multicolumn{1}{r|}{20950}      & \multicolumn{1}{r|}{21811}      & \multicolumn{1}{r}{23501}
    \\ \hline
    
    \multicolumn{13}{l}{k = 32}           
    \\ \hline
    \multicolumn{1}{l|}{Filter+TC}               
    & \multicolumn{1}{r|}{510}      & \multicolumn{1}{r|}{649} &      \multicolumn{1}{r|}{2193} 
    & \multicolumn{1}{r|}{369}      & \multicolumn{1}{r|}{327}      & \multicolumn{1}{r|}{634}      
    & \multicolumn{1}{r|}{5}      & \multicolumn{1}{r|}{14}      & \multicolumn{1}{r|}{541}      
    & \multicolumn{1}{r|}{342}      & \multicolumn{1}{r|}{528}      & \multicolumn{1}{r}{2326}
    \\ \hline
    \multicolumn{1}{l|}{WT\_ALL}
    & \multicolumn{1}{r|}{10358}      & \multicolumn{1}{r|}{13142}      & \multicolumn{1}{r|}{11564}      
    & \multicolumn{1}{r|}{1973}      & \multicolumn{1}{r|}{1696}      & \multicolumn{1}{r|}{1617}      
    & \multicolumn{1}{r|}{496}      & \multicolumn{1}{r|}{517}      & \multicolumn{1}{r|}{498}      
    & \multicolumn{1}{r|}{46232}      & \multicolumn{1}{r|}{47935}      & \multicolumn{1}{r}{48784}
    \\ \hline
            
    \end{tabular}

\end{adjustbox}



}

  \caption{Total query time in milliseconds on all data sets for fixed $m_i = 3$ and gap constraints 
           $C_S = \langle 100,110\rangle$, $C_M = \langle 1000,1100\rangle$, and $C_L = \langle 10000,11000\rangle$.
  }
  \label{tab:gapcons}
\end{center}
\end{table}

\paragraph{Matching Performance for Different Subpattern Lengths.}
In our second experiment, we examine the impact of subpattern lengths on query time, fixing the gap constraint to $C_S = 100,110$.
Table~\ref{tab:subpatlen} shows the results. Larger subpattern lengths tend to result in smaller $\SA$ ranges. Consequently, {\sc SA-scan} 
outperforms {\sc WT} by an even wider margin.
\begin{table}[htb]
\begin{center}

  \let\center\empty
  \let\endcenter\relax
  \centering
  \resizebox{1\width}{!}{




\centering
\begin{adjustbox}{width=\textwidth}

    \begin{tabular}{lllllllllllll}
    \hline
    \multicolumn{1}{l|}{\multirow{2}{*}{Method}} 
        & \multicolumn{3}{l|}{CC} 
        & \multicolumn{3}{l|}{Kernel} 
        & \multicolumn{3}{l|}{Para}
        & \multicolumn{3}{l}{Proteins}
        \\ 
    \cline{2-13} 
    \multicolumn{1}{r|}{} 
    & \multicolumn{1}{c|}{3} & \multicolumn{1}{c|}{5} & \multicolumn{1}{c|}{7} 
    & \multicolumn{1}{c|}{3} & \multicolumn{1}{c|}{5} & \multicolumn{1}{c|}{7} 
    & \multicolumn{1}{c|}{3} & \multicolumn{1}{c|}{5} & \multicolumn{1}{c|}{7}
    & \multicolumn{1}{c|}{3} & \multicolumn{1}{c|}{5} & \multicolumn{1}{c}{7} 
    \\ 
    \hline
    \multicolumn{13}{l}{k = 2}
    \\ 
    \hline
    \multicolumn{1}{l|}{Filter+TC}               
    & \multicolumn{1}{r|}{1391}      & \multicolumn{1}{r|}{478}      & \multicolumn{1}{r|}{321}      
    & \multicolumn{1}{r|}{744}      & \multicolumn{1}{r|}{304}      & \multicolumn{1}{r|}{84}  
    & \multicolumn{1}{r|}{2243}      & \multicolumn{1}{r|}{629}      & \multicolumn{1}{r|}{75} 
   & \multicolumn{1}{r|}{388}      & \multicolumn{1}{r|}{33}      & \multicolumn{1}{r}{24} 
    
    \\ \hline
    \multicolumn{1}{l|}{WT\_ALL}                 
    & \multicolumn{1}{r|}{17354}      & \multicolumn{1}{r|}{4535}      & \multicolumn{1}{r|}{2995}      
    & \multicolumn{1}{r|}{7393}      & \multicolumn{1}{r|}{1725}      & \multicolumn{1}{r|}{94}      
    & \multicolumn{1}{r|}{12372}      & \multicolumn{1}{r|}{9911}      & \multicolumn{1}{r|}{2435}
    & \multicolumn{1}{r|}{8549}      & \multicolumn{1}{r|}{427}      & \multicolumn{1}{r}{172}
    
    \\ \hline
    \multicolumn{13}{l}{k = 4}           
    \\ \hline
    \multicolumn{1}{l|}{Filter+TC}               
    & \multicolumn{1}{r|}{1099}      & \multicolumn{1}{r|}{327}      & \multicolumn{1}{r|}{179}      
    & \multicolumn{1}{r|}{664}      & \multicolumn{1}{r|}{132}      & \multicolumn{1}{r|}{64}      
    & \multicolumn{1}{r|}{1214}      & \multicolumn{1}{r|}{742}      &       \multicolumn{1}{r|}{67}
    & \multicolumn{1}{r|}{370}      & \multicolumn{1}{r|}{32}      & \multicolumn{1}{r}{21}
    \\ \hline
    \multicolumn{1}{l|}{WT\_ALL}                 
    & \multicolumn{1}{r|}{7452}      & \multicolumn{1}{r|}{2030}      & \multicolumn{1}{r|}{512}      
    & \multicolumn{1}{r|}{2201}      & \multicolumn{1}{r|}{145}      & \multicolumn{1}{r|}{31}      
    & \multicolumn{1}{r|}{6554}      & \multicolumn{1}{r|}{10908}      &       \multicolumn{1}{r|}{2414}
    & \multicolumn{1}{r|}{8486}      & \multicolumn{1}{r|}{96}      & \multicolumn{1}{r}{68}
    \\ \hline
    
    \multicolumn{13}{l}{k = 8}           
    \\ \hline
    \multicolumn{1}{l|}{Filter+TC}               
    & \multicolumn{1}{r|}{2185}      & \multicolumn{1}{r|}{227} &      \multicolumn{1}{r|}{100} 
    & \multicolumn{1}{r|}{328}      & \multicolumn{1}{r|}{133}      & \multicolumn{1}{r|}{116}      
    & \multicolumn{1}{r|}{5}      & \multicolumn{1}{r|}{628}      & \multicolumn{1}{r|}{66}      
    & \multicolumn{1}{r|}{354}      & \multicolumn{1}{r|}{28}      & \multicolumn{1}{r}{25}
    \\ \hline
    \multicolumn{1}{l|}{WT\_ALL}
    & \multicolumn{1}{r|}{7919}      & \multicolumn{1}{r|}{1395}      & \multicolumn{1}{r|}{466}      
    & \multicolumn{1}{r|}{1885}      & \multicolumn{1}{r|}{154}      & \multicolumn{1}{r|}{47}      
    & \multicolumn{1}{r|}{237}      & \multicolumn{1}{r|}{16475}      & \multicolumn{1}{r|}{3563}      
    & \multicolumn{1}{r|}{11940}      & \multicolumn{1}{r|}{48}      & \multicolumn{1}{r}{33}
    \\ \hline
    
    \multicolumn{13}{l}{k = 16}           
    \\ \hline
    \multicolumn{1}{l|}{Filter+TC}               
    & \multicolumn{1}{r|}{674}      & \multicolumn{1}{r|}{228} &      \multicolumn{1}{r|}{140} 
    & \multicolumn{1}{r|}{312}      & \multicolumn{1}{r|}{90}      & \multicolumn{1}{r|}{50}      
    & \multicolumn{1}{r|}{5}      & \multicolumn{1}{r|}{675}      & \multicolumn{1}{r|}{65}      
    & \multicolumn{1}{r|}{350}      & \multicolumn{1}{r|}{28}      & \multicolumn{1}{r}{33}
    \\ \hline
    \multicolumn{1}{l|}{WT\_ALL}
    & \multicolumn{1}{r|}{7483}      & \multicolumn{1}{r|}{925}      & \multicolumn{1}{r|}{365}      
    & \multicolumn{1}{r|}{948}      & \multicolumn{1}{r|}{129}      & \multicolumn{1}{r|}{80}      
    & \multicolumn{1}{r|}{279}      & \multicolumn{1}{r|}{29231}      & \multicolumn{1}{r|}{6222}      
    & \multicolumn{1}{r|}{20950}      & \multicolumn{1}{r|}{83}      & \multicolumn{1}{r}{61}
    \\ \hline
    
    \multicolumn{13}{l}{k = 32}           
    \\ \hline
    \multicolumn{1}{l|}{Filter+TC}               
    & \multicolumn{1}{r|}{510}      & \multicolumn{1}{r|}{192} &      \multicolumn{1}{r|}{109} 
    & \multicolumn{1}{r|}{369}      & \multicolumn{1}{r|}{117}      & \multicolumn{1}{r|}{85}      
    & \multicolumn{1}{r|}{5}      & \multicolumn{1}{r|}{563}      & \multicolumn{1}{r|}{65}      
    & \multicolumn{1}{r|}{342}      & \multicolumn{1}{r|}{32}      & \multicolumn{1}{r}{28}
    \\ \hline
    \multicolumn{1}{l|}{WT\_ALL}
    & \multicolumn{1}{r|}{10358}      & \multicolumn{1}{r|}{1464}      & \multicolumn{1}{r|}{548}      
    & \multicolumn{1}{r|}{1973}      & \multicolumn{1}{r|}{274}      & \multicolumn{1}{r|}{167}      
    & \multicolumn{1}{r|}{496}      & \multicolumn{1}{r|}{64780}      & \multicolumn{1}{r|}{13435}      
    & \multicolumn{1}{r|}{46232}      & \multicolumn{1}{r|}{181}      & \multicolumn{1}{r}{131}
    \\ \hline
            
    \end{tabular}

\end{adjustbox}



}

  \caption{Total query time in milliseconds for fixed gap constraint $C_S = \langle 100,110\rangle$ for different  
           subpattern lengths $m_i \in \lbrace 3, 5, 7\rbrace$ and different data sets.
  }
  \label{tab:subpatlen}
\end{center}
\end{table}

\begin{figure}[htb]
  \includegraphics[width=0.9\linewidth]{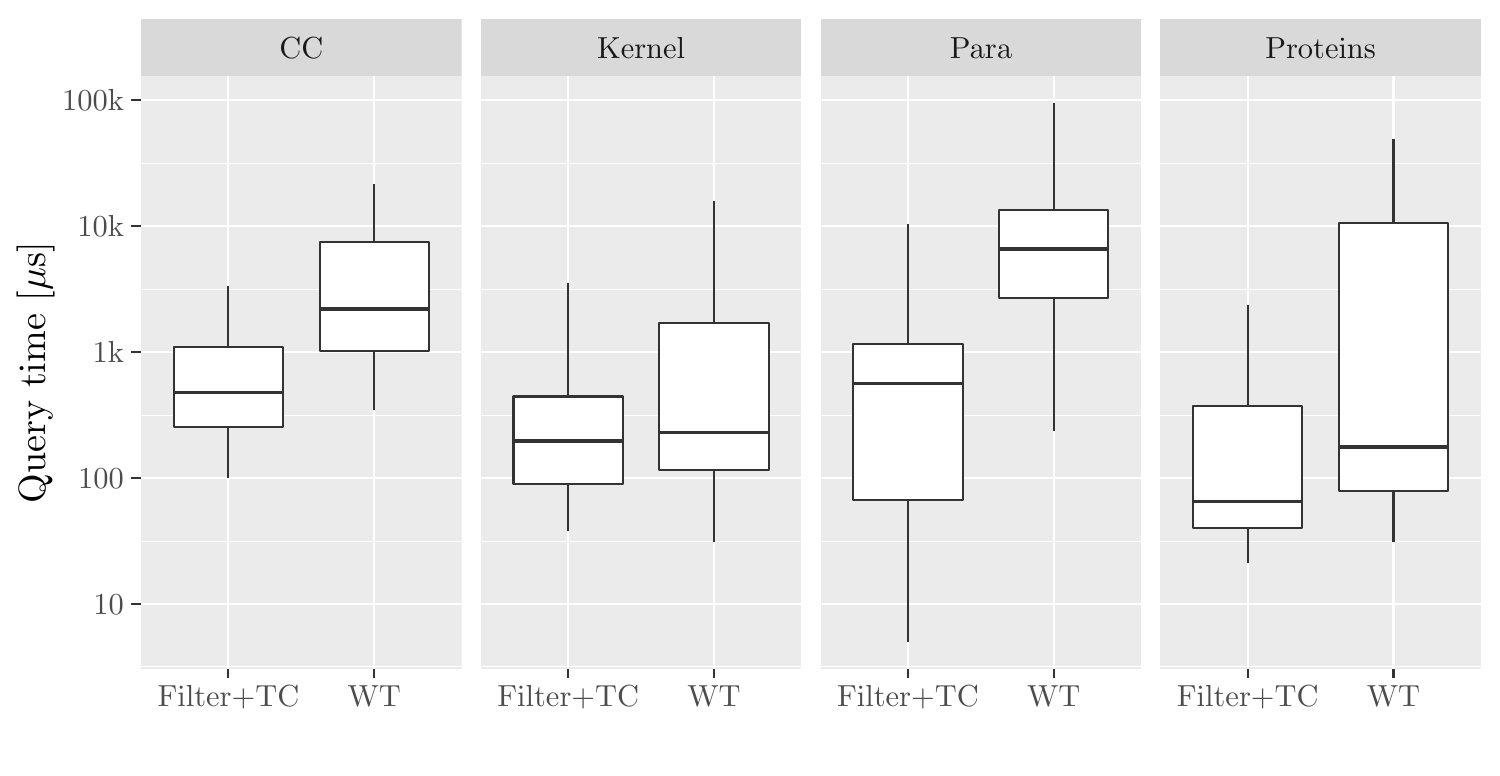}
  \caption{Overall runtime performance of both methods, accumulating the performance for 
           all $m_i \in \lbrace 3, 5, 7\rbrace$ and $C_S$, $C_M$, and $C_L$.
  }
  \label{fig:overall}
\end{figure}

\paragraph{Overall Runtime Performance.}
In a final experiment we explored the whole parameter space (i.e. $k \in \lbrace 2^1, \ldots, 2^5\rbrace$, $m_i \in \lbrace 3, 5, 7\rbrace$, $C \in \lbrace C_S, C_M, C_L\rbrace$). The results are summarized in Figure~\ref{fig:overall}. Overall out {\sc SA-scan}-based method is faster on average than the wavelet-tree-based one, usually by a wide margin.

\section{Concluding Remarks}
\label{sec:conclusion}
We have described a number of simple but highly effective improvements to the {\sc SA-scan} VLG matching algorithm that, 
according to our experiments, elevate it to be the state-of-the-art approach for the indexed version of problem. We believe 
better indexing methods for VLG matching can be found, but that our version of {\sc SA-scan}, which makes judicious use of 
filters, text checking, and subpattern processing order, represents a strong baseline against which the performance of more 
exotic methods should be measured. 

Numerous avenues for continued work on VLG matching exist, perhaps the most interesting of which is to reduce index size.
Currently, {\sc SA-scan} uses $n\log n + n\log\sigma$ bits of space for a text of length $n$ on alphabet $\sigma$ for the 
suffix array and text, respectively (the {\sc WT} approach of Bader et al., uses slightly more). 
Because our methods consist (mostly) of simple scans of SA ranges or scans of the underlying text, they are easily translated 
to make use of recent results on Burrows-Wheeler-based compressed indexes~\cite{GNP18} that allow fast access to elements 
of the suffix array from a compressed representation of it. Via this observation we derive the first compressed indexes for VLG 
matching. These indexes use $O(r\log n)$ bits of space, where $r$ is the number of runs in the Burrows-Wheeler transform, a 
quantity that decreases with text compressibility. On our 2~GiB Kernel data set, for example, the compressed index takes around 
20~MiB in practice, and can still support VLG matching in times competitive with the indexes of Bader et al. We plan to explore 
this in more depth in future work.

\paragraph{Acknowledgements.}
Our thanks go to Tania Starikovskaya for suggesting the problem of indexing for regular-expression
matching to us.
We also thank Matthias Petri and Simon Gog for prompt answers to questions about 
their article and code and the anonymous reviewers for helpful comments. This work was funded by the Academy of Finland via grant 319454 and by EU’s Horizon 
2020 research and innovation programme under Marie Sk{\l}odowska-Curie grant agreement No 690941 (BIRDS).

\bibliographystyle{splncs03}
\bibliography{sofsem}

\begin{thebibliography}{10}
\providecommand{\url}[1]{\texttt{#1}}
\providecommand{\urlprefix}{URL }

\bibitem{BGP16}
Bader, J., Gog, S., Petri, M.: Practical variable length gap pattern matching.
  In: Goldberg, A.V., Kulikov, A.S. (eds.) Proc. SEA. pp. 1--16. LNCS 9685
  (2016)

\bibitem{BFC08}
Bille, P., Farach{-}Colton, M.: Fast and compact regular expression matching.
  Theor. Comput. Sci.  409(3),  486--496 (2008)

\bibitem{BG14}
Bille, P., G{\o}rtz, I.L.: Substring range reporting. Algorithmica  69(2),
  384--396 (2014)

\bibitem{BGVW12}
Bille, P., G{\o}rtz, I.L., Vildh{\o}j, H.W., Wind, D.K.: String matching with
  variable length gaps. Theor. Comput. Sci.  443,  25--34 (2012)

\bibitem{BT10}
Bille, P., Thorup, M.: Regular expression matching with multi-strings and
  intervals. In: Proc. SODA. pp. 1297--1308. {ACM-SIAM} (2010)

\bibitem{C12}
Cox, R.: Regular expression matching with a trigram index or how {G}oogle code
  search worked (2012), \url{https://swtch.com/~rsc/regexp/regexp4.html}

\bibitem{CIR98}
Crawford, T., Iliopoulos, C.S., Raman, R.: String matching techniques for
  musical similarity and melodic recognition. Computing in Musicology  11,
  73--100 (1998)

\bibitem{CIMRTT02}
Crochemore, M., Iliopoulos, C.S., Makris, C., Rytter, W., Tsakalidis, A.K.,
  Tsichlas, T.: Approximate string matching with gaps. N. J. Comput.  9(1),
  54--65 (2002)

\bibitem{FG08}
Fredriksson, K., Grabowski, S.: Efficient algorithms for pattern matching with
  general gaps, character classes, and transposition invariance. Inf. Retr.
  11(4),  335--357 (2008)

\bibitem{GNP18}
Gagie, T., Navarro, G., Prezza, N.: Optimal-time text indexing in {BWT}-runs
  bounded space. In: Proc. SODA. pp. 1459--1477. {ACM-SIAM} (2018)

\bibitem{GGV03}
Grossi, R., Gupta, A., Vitter, J.: High-order entropy-compressed text indexes.
  In: Proc. SODA. pp. 841--850. ACM-SIAM (2003)

\bibitem{HSSSS11}
Haapasalo, T., Silvasti, P., Sippu, S., Soisalon{-}Soininen, E.: Online
  dictionary matching with variable-length gaps. In: Proc. SEA. pp. 76--87.
  LNCS 6630 (2011)

\bibitem{KMP}
Knuth, D., Morris, J.H., Pratt, V.: Fast pattern matching in strings. SIAM
  Journal on Computing  6(2),  323--350 (1977)

\bibitem{L11}
Lewenstein, M.: Indexing with gaps. In: Proc. SPIRE. pp. 135--143. LNCS 7024,
  Springer (2011)

\bibitem{L07}
Lopez, A.: Hierarchical phrase-based translation with suffix arrays. In: Proc.
  EMNLP-CoNLL 2007. pp. 976--985. {ACL} (2007)

\bibitem{MM93}
Manber, U., Myers, G.: Suffix arrays: a new method for on-line string searches.
  SIAM J. Computing  22(5),  935--948 (1993)

\bibitem{MC05}
Metzler, D., Croft, W.B.: A markov random field model for term dependencies.
  In: Proc. {SIGIR}. pp. 472--479. {ACM} (2005)

\bibitem{MPVZ05}
Morgante, M., Policriti, A., Vitacolonna, N., Zuccolo, A.: Structured motifs
  search. Journal of Computational Biology  12(8),  1065--1082 (2005)

\bibitem{N13}
Navarro, G.: Wavelet trees for all. Journal of Discrete Algorithms  25,  2--20
  (2014)

\bibitem{P14}
Pissis, S.P.: {M}o{T}e{X}-{II}: structured {M}o{T}if e{X}traction from
  large-scale datasets. BMC Bioinformatics  15(235),  1--12 (2014)

\bibitem{SSSS15}
Saikkonen, R., Sippu, S., Soisalon{-}Soininen, E.: Experimental analysis of an
  online dictionary matching algorithm for regular expressions with gaps. In:
  Proc. SEA. pp. 327--338. LNCS 9125, Springer (2015)

\bibitem{TTHW07}
Turpin, A., Tsegay, Y., Hawking, D., Williams, H.E.: Fast generation of result
  snippets in web search. In: Proc. {SIGIR} 2007. pp. 127--134. {ACM} (2007)

\end{thebibliography}

\end{document}